\documentclass[aps,prb, twocolumn] {revtex4}
\usepackage{epsfig}

\begin{document}
\headsep 2.5cm

\title{Comment on "Locally self-consistent embedding approach for disordered electronic systems". }
\author{Rostam Moradian$^{1,2}$, Sina Moradian$^3$, Rouhollah Gholami$^4$}

\affiliation{$^{1}$Physics Department, Faculty of Science, Razi
University, Kermanshah, Iran\\
$^{2}$Nano Science and Nano Technology Research Center, Razi University, Kermanshah, Iran\\
$^3$Department of Electrical and Computer Engineering, University of Central Florida, Orlando, Florida, USA\\
$^{4}$Physics Department, Faculty of Science, Ilam
University, Ilam, Iran.}

\begin{abstract}
 We comment on article by Yi Zhang , Hanna Terletska, Ka-Ming Tam, Yang Wang, Markus Eisenbach,  Liviu Chioncel, and Mark Jarrell [Phys. Rev. B {\bf 100}, 054205 (2019)]\cite{Zhang} in which to study substitution disordered systems, they presented an embedding scheme for the locally self-consistent method. Here we show that their methods is a truncated case of our super-cell approximation, achieved by neglecting super-cell wave vectors dependence on self-energy $\Sigma_{sc}({\bf K}_{n},E)$ and replacing them by a local on-site self-energy, $\Sigma_{sc}({\bf K}_{n},E)=\Sigma_{sc}(L,L,E)$ in our articles\cite{Moradian01, Moradian02, Moradian03}. Also their real and k-space self-energies in the limit of the number of super-cell sites, $N_{c}$, approaching the number of lattice sites, N, do not recover exact self-energies $\Sigma(l, l', E)$ and  $\Sigma({\bf k}, E)$. For highlighting advantages of our methods with respect to other approximations such as dynamical cluster approximation (DCA)\cite{Jarrell} in capturing electron localization, we apply our real space super-cell approximation (SCA),  and super-cell local self-energy approximation (SCLSA) to one and two dimensional substitution disorder alloy systems. Our electron localization probability calculations for these systems determine non zero values that indicate electrons localization.
\end{abstract}
 \maketitle

In the super-cell approximation, real space disorder systems are divided into super-cells with original lattice symmetries and $N_{c}$ sites. In general, impurity configurations of super-cells are not identical, so no periodic boundary condition between super-cells.We proved that taking impurity configurations averaging and neglecting k-space self-energy contributions of two sites in different super-cells leads to k-space super-cell self-energies, $\{\Sigma({\bf K}_{1}, E),...,\Sigma({\bf K}_{N_{c}}, E)\}$, and also periodicity of real space self-energy, $\Sigma(I,J, E)$, with respect to super-cell lengths\cite{Moradian01, Moradian02, Moradian03}. In our methods relation between real space and K-space cavity Green functions defined by ${\mathcal G}(I, J; E)=\frac{1}{Nc}\sum_{\bf K}e^{i{\bf K}.{\bf r}_{IJ}}{\mathcal G}({\bf K}; E)$\cite{Moradian01} that is different than DCA definition ${\mathcal G}(I, J; E)=\sum_{\bf K}e^{i{\bf K}.{\bf r}_{IJ}}{\mathcal G}({\bf K}; E)$\cite{Jarrell}.  In the limit of $Nc\longrightarrow N$, our K-space cavity Green functions converts to clean k-space Green function, $G^{0}({\bf k}; E)=\frac{1}{E+i\eta +\epsilon_{\bf k}}$, $lim_{Nc\longrightarrow N}{\mathcal G}({\bf K}; E)=G^{0}({\bf k}; E)$, and real space cavity Green functions converts to real space clean Green function $lim_{Nc\longrightarrow N}{\mathcal G}(I, J; E)=G^{0}(i,j; E)=\frac{1}{N}\sum_{i,j} e^{i{\bf k}.{\bf r}_{ij}}G^{0}({\bf k}; E)$ while for the DCA real space cavity Green function converts to $lim_{Nc\longrightarrow N}{\mathcal G}(I, J; E)=\sum_{i,j} e^{i{\bf k}.{\bf r}_{ij}}G^{0}({\bf k}; E)$.  Note that our beyond supper-cell approximation solves the k-space self-energy discontinuity problem of DCA and super-cell approximation.

Yi Zhang et al. \cite{Zhang} by neglecting nonlocal real space self-energies, $\Sigma_{sc}(L,L', E)=\Sigma_{sc}(L,L, E)\delta_{LL'}$, without  derivation introduced relation between local self-energy $\Sigma_{sc}(L,L, E)$ and average super-cell Green function as $\bar{G}({\bf K}_{n}, E)=\frac{N_{c}}{N}\sum_{{\bf k}\in nth\; grain}\frac{1}{E+i\eta+\epsilon_{\bf k}-\Sigma_{sc}(L,L, E)}$, that is Eq.3 of Ref. 1.  One way for checking correctness of a method is that its equations in specific cases convert  to lower approximation equations or exact equations. For disorder systems lower approximation case of cluster approximations is single site coherent potential approximation (CPA) where $Nc=1$ and exact case is cluster with whole lattice that is $Nc=N$. First we check their coarse grained Green function $\bar{G}({\bf K}_{n}, E)=\frac{N_{c}}{N}\sum_{{\bf k}\in nth\; grain}\frac{1}{E+i\eta+\epsilon_{\bf k}-\Sigma(L,L, E)}$ in the case $Nc=1$ and $K_{1}=0$. Their local average Green function in this case becomes
\begin{eqnarray}
\bar{G}(0, E)=\frac{1}{N}\sum_{{\bf k}\in FBZ}\frac{1}{E+i\eta+\epsilon_{\bf k}-\Sigma(L,L, E)}
\label{eq:k-av-green-cpa}
\end{eqnarray}
 which recovers CPA average Green function  successfully. For another limited case of $Nc=N$, as in the DCA \cite{Jarrell} and our methods \cite{Moradian01} shown,  the cluster $K_{n}$ converts to ${\bf k}$ in the whole first Brillouin zone(FBZ) and k-space coarse grained Green function converts to exact Green function $\bar{G}({\bf k}, E)=\frac{1}{E+i\eta+\epsilon_{\bf k}-\Sigma({\bf k}, E)}$. For this case, $N_{c}\rightarrow N$, Zhang et al. k-space and real space average Green function becomes  
\begin{eqnarray}
\bar{G}({\bf k}, E)&=&\frac{1}{E+i\eta+\epsilon_{\bf k}-\Sigma(L,L, E)},\nonumber\\\bar{G}(i,j, E)&=&\frac{1}{N}\sum_{{\bf k}\in FBZ}\frac{e^{i{\bf k}.{\bf r}_{ij}}}{E+i\eta+\epsilon_{\bf k}-\Sigma(L,L, E)} \nonumber\\
\label{eq:k-av-green}
\end{eqnarray}
 which does not recover k-space and real space exact average Green functions. So this method is not successful for clusters with high number of sites, $Nc\rightarrow N$. The big weakness point of Yi Zhang et al. \cite{Zhang} is neglecting  self energy k-dependent which is successful just for high dimension systems as proved by  R. Vlaming and D. Vollhardt\cite{Vlaming}, and D. W. Metzner and D. Vollhardt\cite{Metzner}. 

 By comparing our cavity Green function Eq.13 of Ref.4 with Eqs.6 and 7 of Ref.1 we see that they are same except their real space self-energy is site diagonal. Another problem of Zhang et al. formalism in their algorithm , Fig.3 of Ref. 1, is that in Eq.6 without any acceptable reason to obtain Eq.11 of Ref. 1 they substituted super-cell impurity average Green function $\overline{{\bf G}^{LIS}}(E)_{II}$  by  a typical average Green function $\overline{{\bf G}^{LIS}}_{typ}(E)_{II}$ defined by
 \begin{eqnarray}
 \bar{G}_{typ}( E)_{II}&=&\frac{1}{Nc}\sum^{Nc}_{I=1}\left(\overline{\frac{{\bf G}^{LIS}(E)_{II}}{-\frac{1}{\pi}Im {\bf G}^{LIS}(E)_{II}}}\right)\nonumber\\&\times&exp{\left(\frac{1}{Nc}\sum^{Nc}_{I=1}\overline{ln (-\frac{1}{\pi}Im {\bf G}^{LIS}(E)_{II})}\right)}.
 \label{eq:typ-av-green2}
 \end{eqnarray} 
 Also in their algorithm after calculation of ${\mathcal G}^{-1}$ they should calculate ${\bf G}^{LIS}(E, \varepsilon_{I})_{II}$ by using Eq.7 of Ref.1 then calculate $\overline{{\bf G}^{LIS}}(E)_{II}$ and $\overline{{\bf G}^{LIS}}_{typ}(E)_{II}$. Possibly they missed this. It should emphasize that Eq.11 of Ref. 1 in the limit of large super-cell doesn't converts to Dyson equation for whole average system.

 As a case study of the implementation of our method we investigate electron localization in a binary alloy which is key of Anderson metal-insulator phase transition. The probability of 1 and 2 atoms type at $j$th site  is given by $p_{j}=c\delta(\varepsilon_{j}-\varepsilon_{1})+(1-c)\delta(\varepsilon_{j}-\varepsilon_{2})$ in which $\varepsilon_{1}=\delta$ and $\varepsilon_{2}=-\delta$. For a super-cell with $N_{c}$ sites probability of each super-cell impurity configuration is $\Pi^{Nc}_{j=1}p_{j}$.  We calculated electron localization probability for a one and square two dimensional disorder alloy lattices with impurity strength $\delta=4t$ and impurity concentration $c=0.5$, chemical potential $\mu=0$ and band filling $n=1$. For these systems, we use Hamiltonian Eq.1 of Ref.4 in the nearest neighbor approximation with hopping integral $t$.  The electron localization probability at one of the super-cell sites, $L$, defined by\cite{Thouless, McKane} 
\begin{eqnarray}
P(\infty)&=&lim_{t\rightarrow\infty} \overline{|G^{im}(L,L;t)|^{2}}\nonumber\\&=&lim_{\eta\rightarrow 0} \frac{\eta}{\pi}\int dE\overline{|G^{im}(L,L;E+i\eta)|^{2}}
\label{eq:localization}
\end{eqnarray}
where over line means impurity configurations average. To obtain electron localization probability, $P(\eta\rightarrow 0)$, we should calculate $P(\eta)=\frac{\eta}{\pi}\int dE\overline{|G^{im}(L,L;E+i\eta)|^{2}}$ in term of $\eta$ then extrapolate it to $\eta=0$. The algorithm for calculation of super-cell approximation impurity Green function $G^{im}_{sc}(L,L';E+i\eta)$ is as follows,\\
1- A guess for real space and K-space self-energies $\Sigma_{sc}({\bf K}_{ n}; E)$ usually zero.

2- By inserting $\Sigma_{sc}({\bf K}_{ n}; E)$ in $\bar{G}({\bf K}_{ n};E)=\frac{N_{c}}{N}\sum_{{\bf k}\in\;nth\;grain}(G^{-1}_{0}({\bf k};E)-\Sigma_{sc}({\bf K}_{ n};E))^{-1}$,  calculate the average K-space Green functions, $\bar{G}({\bf K}_{n};E)$ .

3- Calculate K-space cavity Green function from $\mathcal{G}({\bf K}_{ n};E)=({\bar G}^{-1}({\bf K}_{ n};E)+\Sigma_{sc}({\bf K}_{n};E))^{-1}$.

4- Fourier transform of K-space $\mathcal{G}({\bf K}_{ n};E)$ to obtained real space  cavity Green function ${\mathcal G}(L,L'; E)=\frac{1}{N_{c}}\sum_{{\bf K}_{n}}e^{i{\bf K}_{n}.{\bf r}_{LL'}}\mathcal{G}({\bf K}_{ n};E)$.

5- Calculate real space impurity Green function matrix from $G^{imp}=({\mathcal G}^{-1}-\mathbf{\varepsilon})^{-1}$ .

6- Calculate electron localization probability from $P_{sc}(\eta)=\frac{\eta}{\pi}\int dE\overline{|G^{im}_{sc}(L,L;E+i\eta)|^{2}}$.

7- Calculate super-cell impurity average Green function matrix, $\bar{ G}_{sc}(L,L'; E)=\overline{(\mathcal{G}^{-1}-\mathbf{\varepsilon})^{-1}}_{LL'}$, by taking an average over all possible impurity configurations.

8- Inverse Fourier transform of new average Green function to calculate ${\bar G}({\bf K}_{n}; E)=\frac{1}{N_{c}}\sum_{LL'}e^{-i{\bf K}_{n}.{\bf r}_{LL'}}{\bar G}_{sc}(L,L'; E)$.

9- Calculate new K-space self-energies from $\Sigma_{sc}({\bf K}_{n}; E)={{\mathcal G}({\bf K}_{ n}; E)}^{-1}-{{\bar G}({\bf K}_{ n};E)}^{-1}$. Then calculate average band filling $n$.

10- Return to 1 and repeat the whole process until convergence of $\Sigma_{sc}({\bf K}_{n}; E)$, average band filling, $n$, and electron localization probability, $P_{sc}(\eta)$, simultaneously. 

Note that our cavity Green function definition is different from DCA. To explore this difference we study the relation between {\bf K}-space cavity Green function, ${\mathcal G}({\bf K}; E)$, and real space ${\mathcal G}(I, J; E)$ in the DCA and based on our method. In the DCA is ${\mathcal G}(I, J; E)=\sum_{\bf K}e^{i{\bf K}.{\bf r}_{IJ}}{\mathcal G}({\bf K}; E)$\cite{Jarrell} while in the super-cell and beyond super-cell approximations is ${\mathcal G}(I, J; E)=\frac{1}{Nc}\sum_{\bf K}e^{i{\bf K}.{\bf r}_{IJ}}{\mathcal G}({\bf K}; E)$\cite{Moradian01} where in the limit of $Nc\longrightarrow N$ our cavity Green functions converts to clean k-space and real space Green functions.

\begin{figure}
	\centerline{\epsfig{file=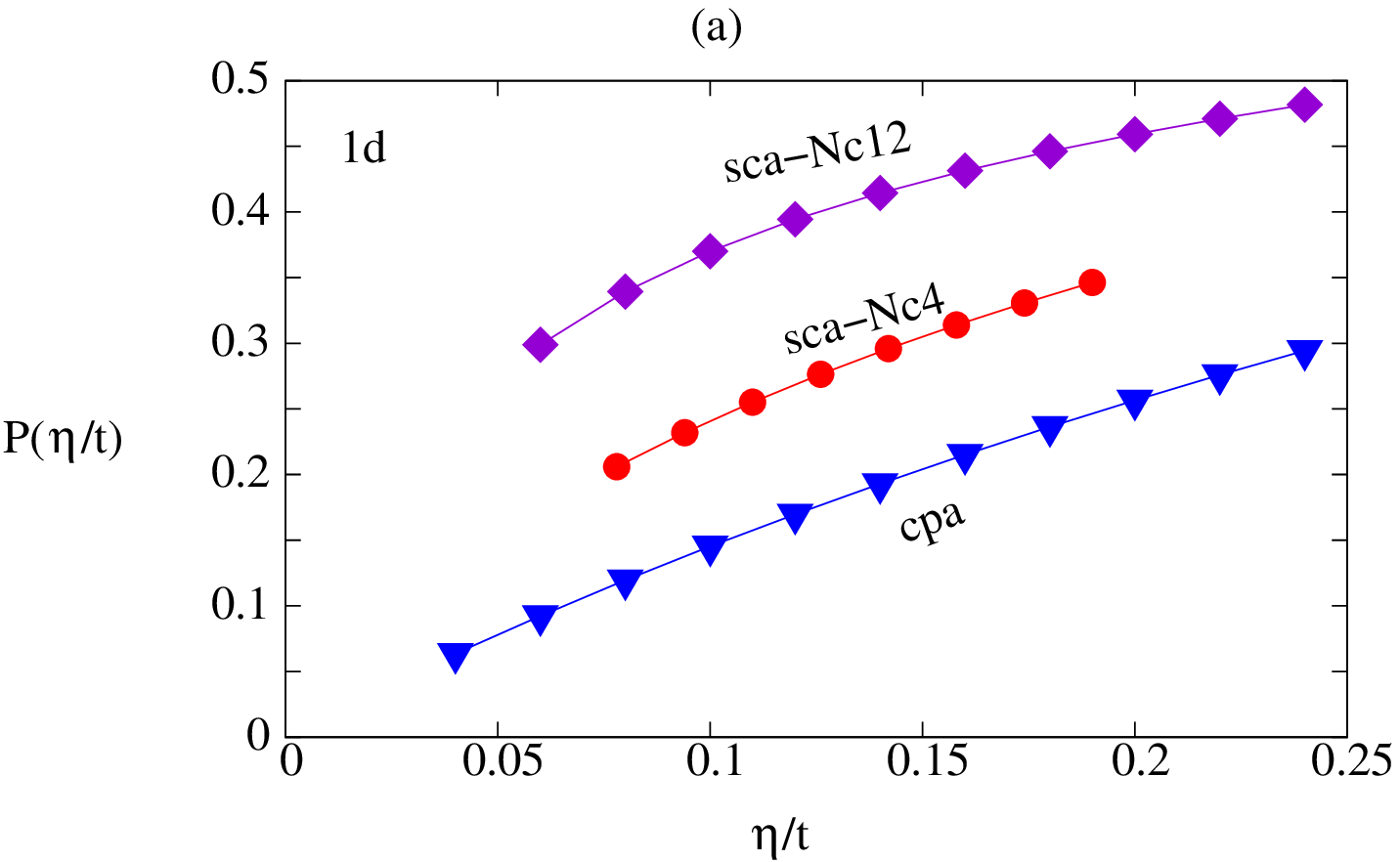 ,width=8.0cm,angle=0}}
\centerline{\epsfig{file=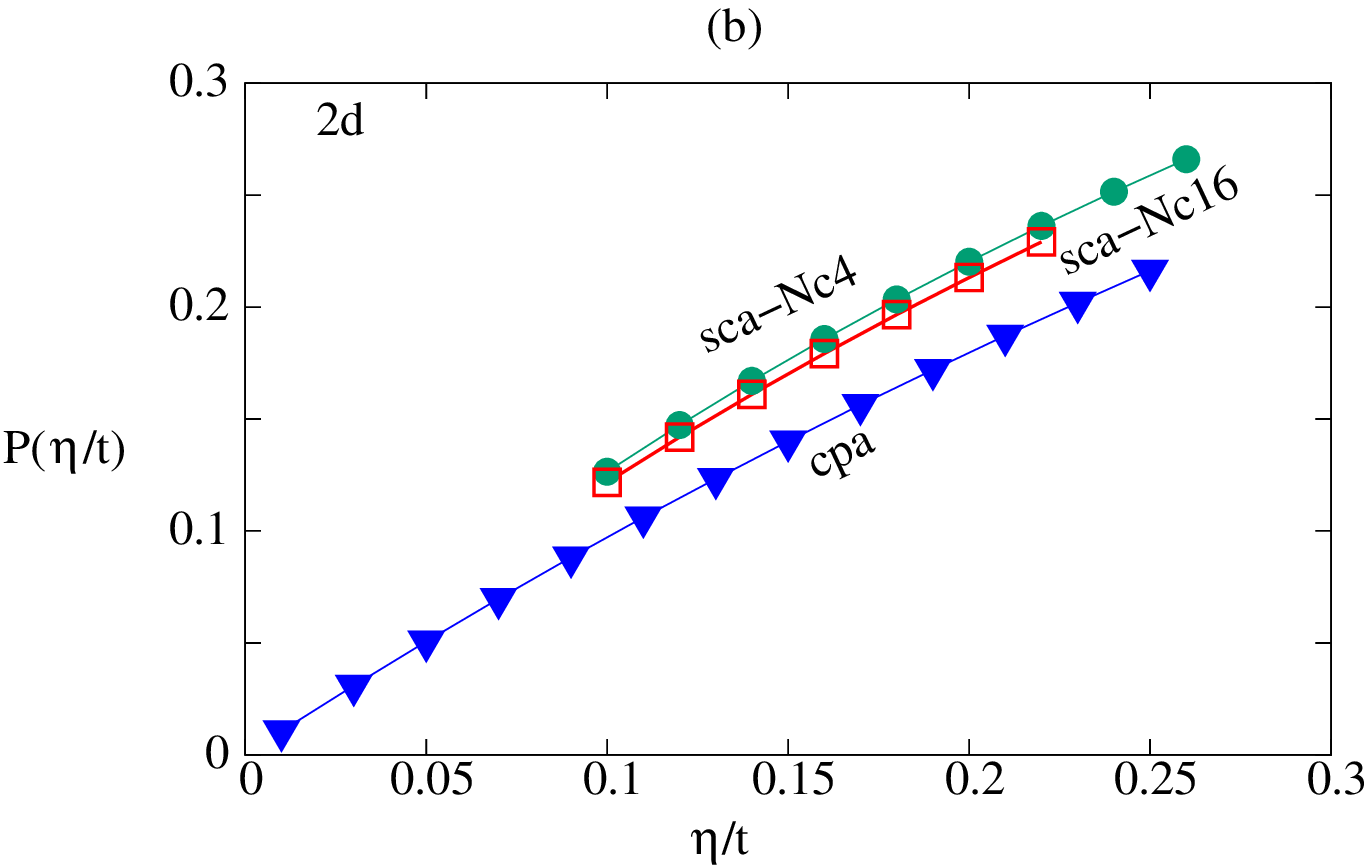 ,width=8.0cm,angle=0}}	
	\caption{(Color on line) Calculated electrons localization probabilities in the super-cell approximation (SCA), (a) for  one-dimensional alloy lattice for different super-cell sizes of $N_{c}=12$, $N_{c}=4$ and CPA $N_{c}=1$, (b) for square two-dimensional alloy lattice for different supercell sizes o $N_{c}=16$, $N_{c}=4$ and CPA $N_{c}=1$. Electron localization probability, $P(\eta\rightarrow 0)$, is zero for CPA but in the cases of $N_{c}>1$ it is finite which means electrons are localized.   }   
	\label{figure:localization-1d-4t-nc12-nc4-nc1-c0.5-n1-c0.5-eta-p} 
\end{figure}

Fig.\ref{figure:localization-1d-4t-nc12-nc4-nc1-c0.5-n1-c0.5-eta-p} shows $P(\eta)$ in terms of $\eta$ for, (a) a one-dimensional alloy in the super-cell approximation for $N_{c}=12$, $N_{c}=4$ and $N_{c}=1$. $P(\eta\rightarrow 0)$ for CPA extrapolates to zero but for $N_{c}=12$ and  $N_{c}=4$ it extrapolates to finite values showing electron localization and (b) a square two-dimensional alloy in the super-cell approximation for $N_{c}=16$, $N_{c}=4$ and $N_{c}=1$. $P(\eta\rightarrow 0)$ for CPA extrapolates to zero but for $N_{c}=16$ and $N_{c}=4$ extrapolate to non zero values that shows electron localization.

The algorithm for calculation of beyond super-cell approximation impurity Green function, ${\bar G}_{bsc}(l,l';E+i\eta)$, is as follows,\\
1- Calculate new super-cell self-energy  $\Sigma_{sc}({\bf K}_{n}; E)$ from super-cell approximation.

2- Calculate beyond super-cell approximation self-energy from $\Sigma_{bsc}({\bf k}; E)=\frac{1}{N^{2}_{c}}\sum_{LL'}\sum_{{\bf K}_{n}}\Sigma_{sc}({\bf K}_{n}; E) e^{i({\bf k}-{\bf K}_{n}).{\bf r}_{LL'}}(1-\frac{1}{N_{c}})$.

3-Calculate beyond super-cell approximation average Green function in the k-space from ${\bar G}_{bsc}({\bf k}, E)=({G^0({\bf k}; E)}^{-1}-\Sigma_{bsc}({\bf k}; E))^{-1}$

3-Fourier transform obtained k-space, self-energies and average Green functions to real space from $\Sigma_{bsc}(l, l'; E)=\frac{1}{N}\sum_{\bf k}\Sigma_{bsc}({\bf k}; E) e^{-i{\bf k}.{\bf r}_{ll'}}$ and ${\bar G}_{bsc}(l, l'; E)=\frac{1}{N}\sum_{\bf k}{\bar G}_{bsc}({\bf k}; E) e^{-i{\bf k}.{\bf r}_{ll'}}$.


In the super-cell local self-energy approximation (SCLSA), all real space super-cell self-energies replace with a local self-energy $\Sigma_{sc}({\bf K}_{n}; E)=\Sigma_{sc}(L,L; E)$ and its algorithm is same as super-cell approximation (SCA).

\begin{figure}
	\centerline{\epsfig{file=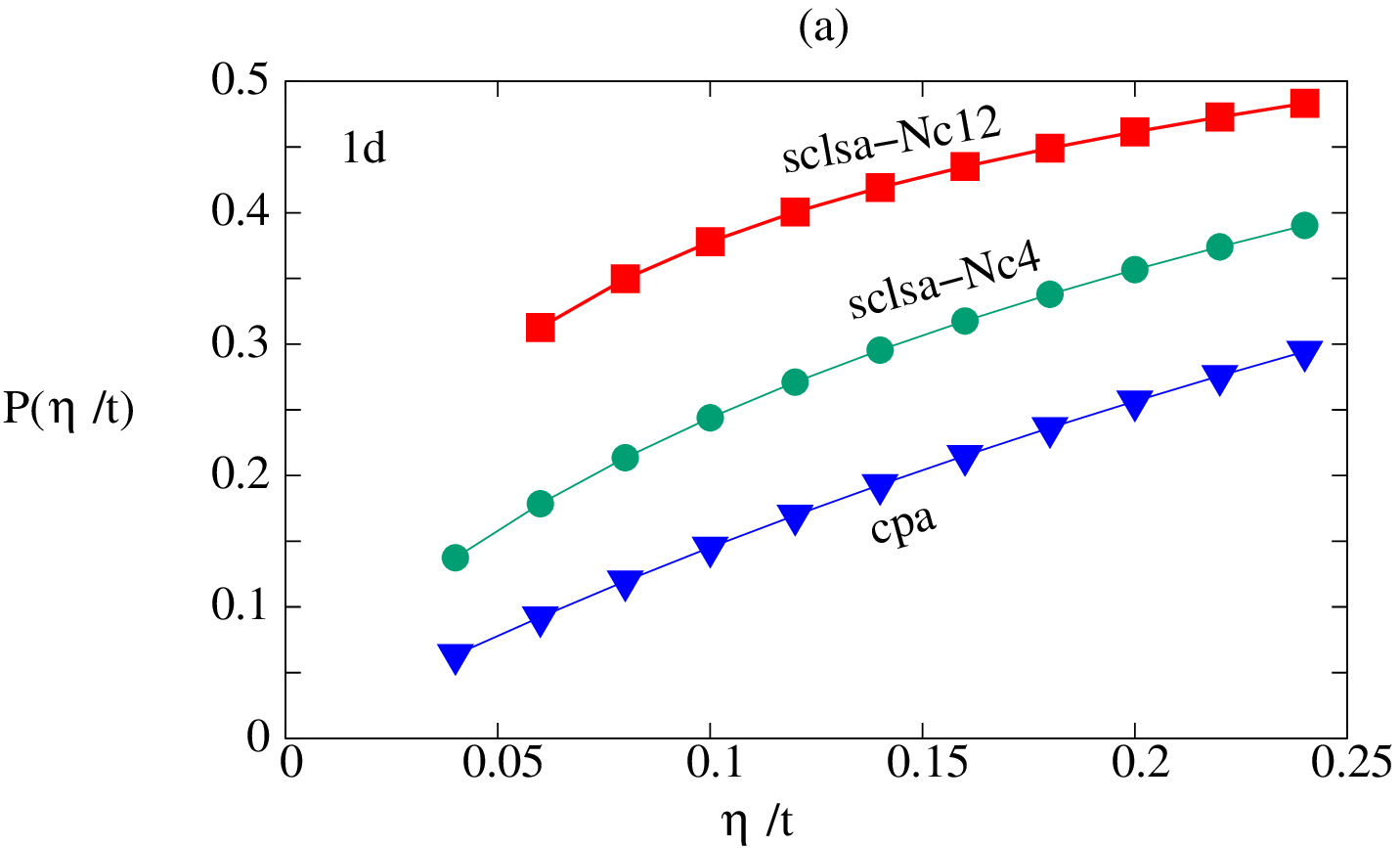 ,width=8.0cm,angle=0}}
    \centerline{\epsfig{file=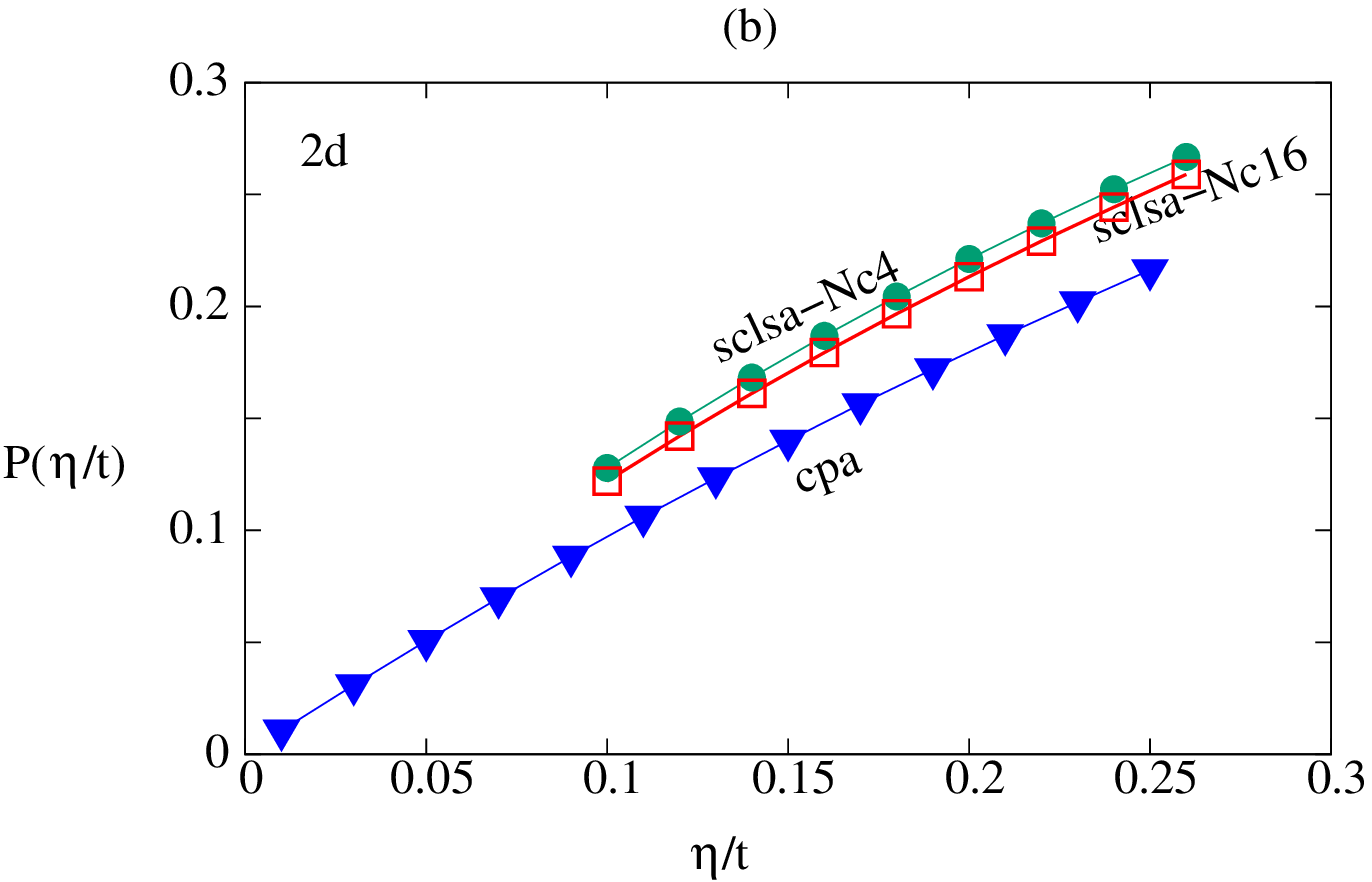 ,width=8.0cm,angle=0}}	
	\caption{(Color on line) Shows electron localization probability in the super cell local self energy approximation (sclsa) for, (a) a one dimensional alloy lattice for different super-cell sizes of $N_{c}=12$, $N_{c}=4$ and CPA $N_{c}=1$, (b) for a square two dimensional alloy lattice for different super-cell sizes of $N_{c}=16$, $N_{c}=4$ and CPA $N_{c}=1$. Electron localization probability $P(\eta\rightarrow 0)$ is zero for CPA but for $N_{c}>1$ it is finite. }   
	\label{figure:localization-1d-4t-nc12-nc4-cpa-c0.5-n1-c0.5-eta-pscpa} 
\end{figure}

Fig.\ref{figure:localization-1d-4t-nc12-nc4-cpa-c0.5-n1-c0.5-eta-pscpa} illustrates electron localization probability for, (a) a one-dimensional alloy lattice in the super-cell local self-energy approximation (sclsa) for different super-cell sizes $N_{c}=12$, $N_{c}=4$ and CPA $N_{c}=1$ and (b) a square two-dimensional lattice for different super-cell sizes $N_{c}=16$, $N_{c}=4$ and CPA $N_{c}=1$. $P(\eta\rightarrow 0)$ extrapolate to finite values for super-cell sizes $N_{c}> 1$, but it is lower than super-cell and beyond super-cell approximations.

To confirm observation of electron localization in the super-cell approximation at high impurity strengths we used localization order parameter method called typical density of states. After performing super-cell self-consistency processes,  we inserted obtained super-cell approximation impurity Green function in the following Dobrosavljevic et al. localization order parameter relation\cite{Dobrosavljević}
\begin{eqnarray}
\bar{G}_{typ}( E)_{II}=exp{\left(\frac{1}{Nc}\sum^{Nc}_{I=1}\overline{ln (-\frac{1}{\pi}Im {\bf G}^{im}_{sc}(E)_{II})}\right)}.
\label{eq:typ-av-green-D}
\end{eqnarray}
where ${\bf G}^{im}_{sc}(E)$ is super-cell impurity Green function matrix and ${\bf G}^{im}_{sc}(E)_{II}$ is its component at $I$th site in the super-cell.

\begin{figure}
	\centerline{\epsfig{file=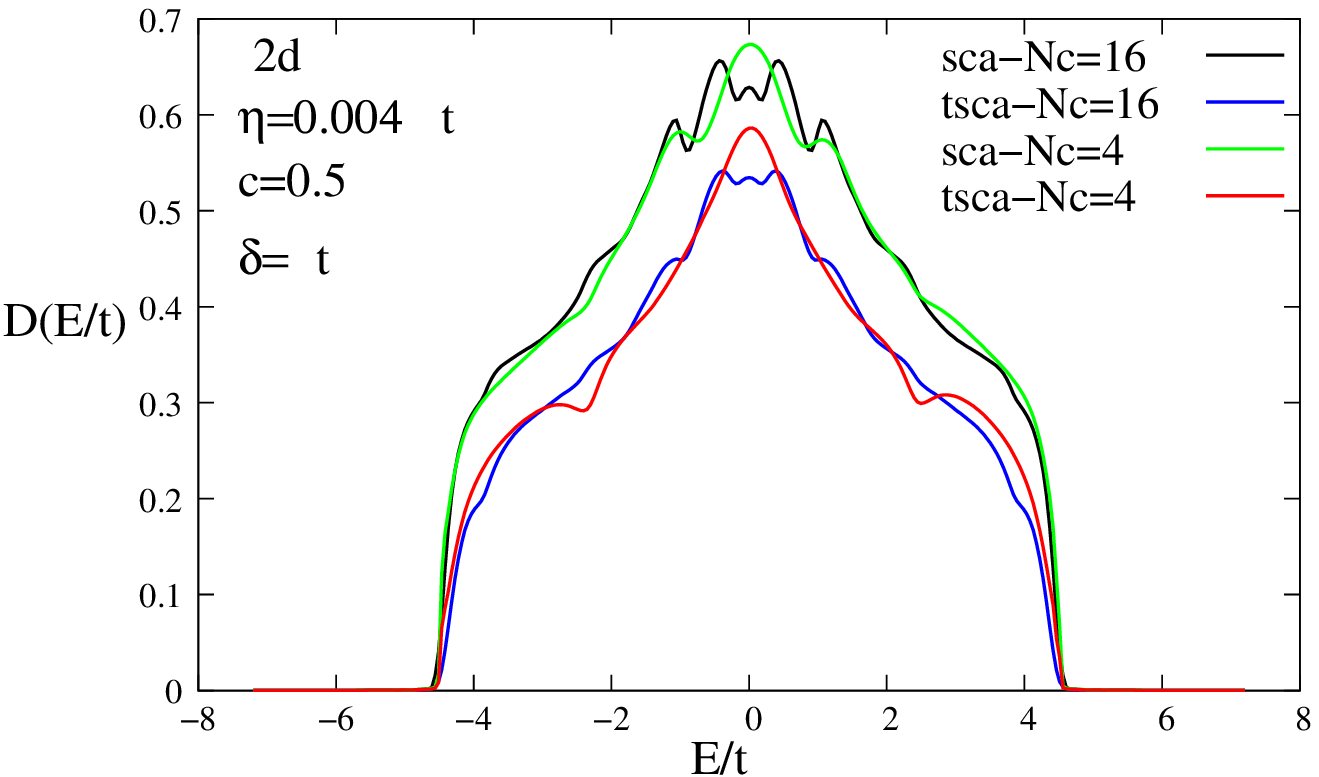  ,width=8.0cm,angle=0}}		
	\centerline{\epsfig{file=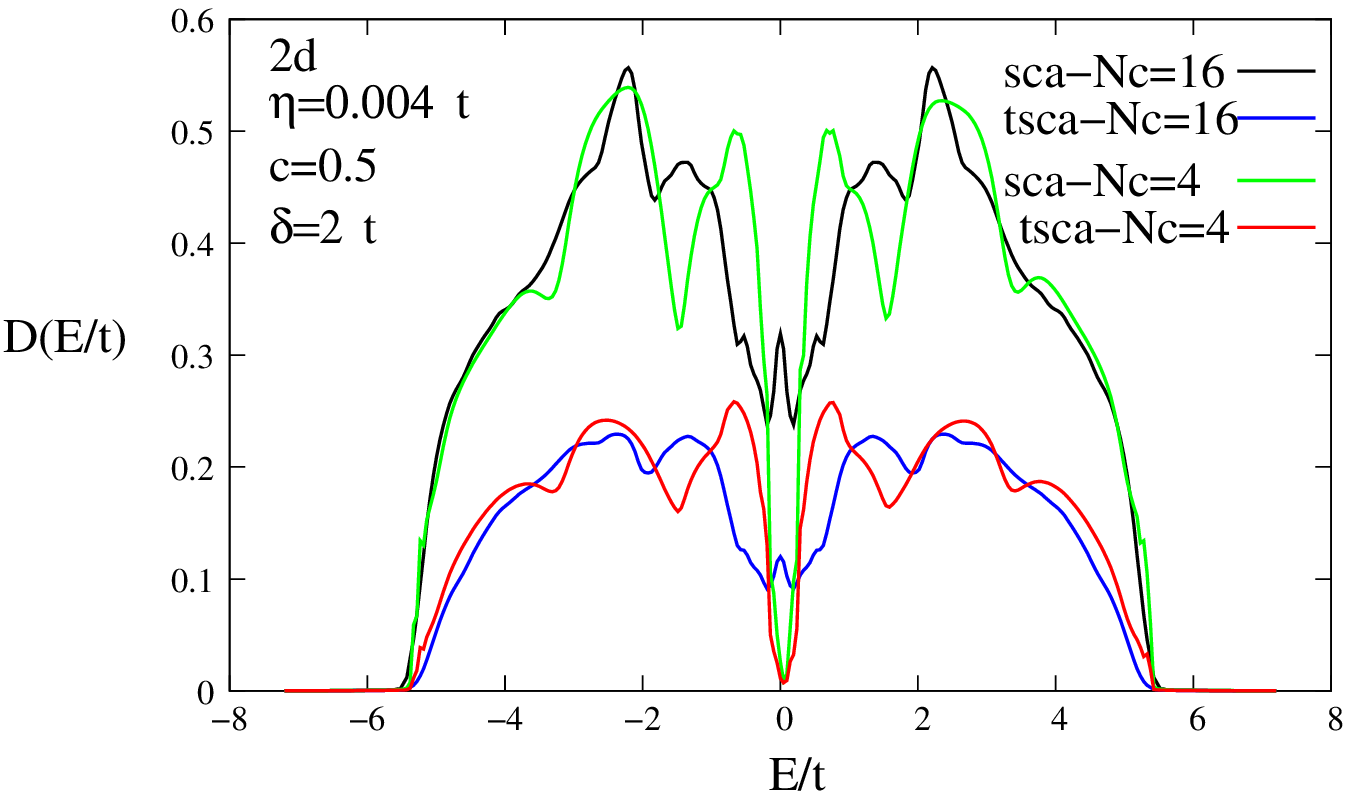  ,width=8.0cm,angle=0}}	
\centerline{\epsfig{file=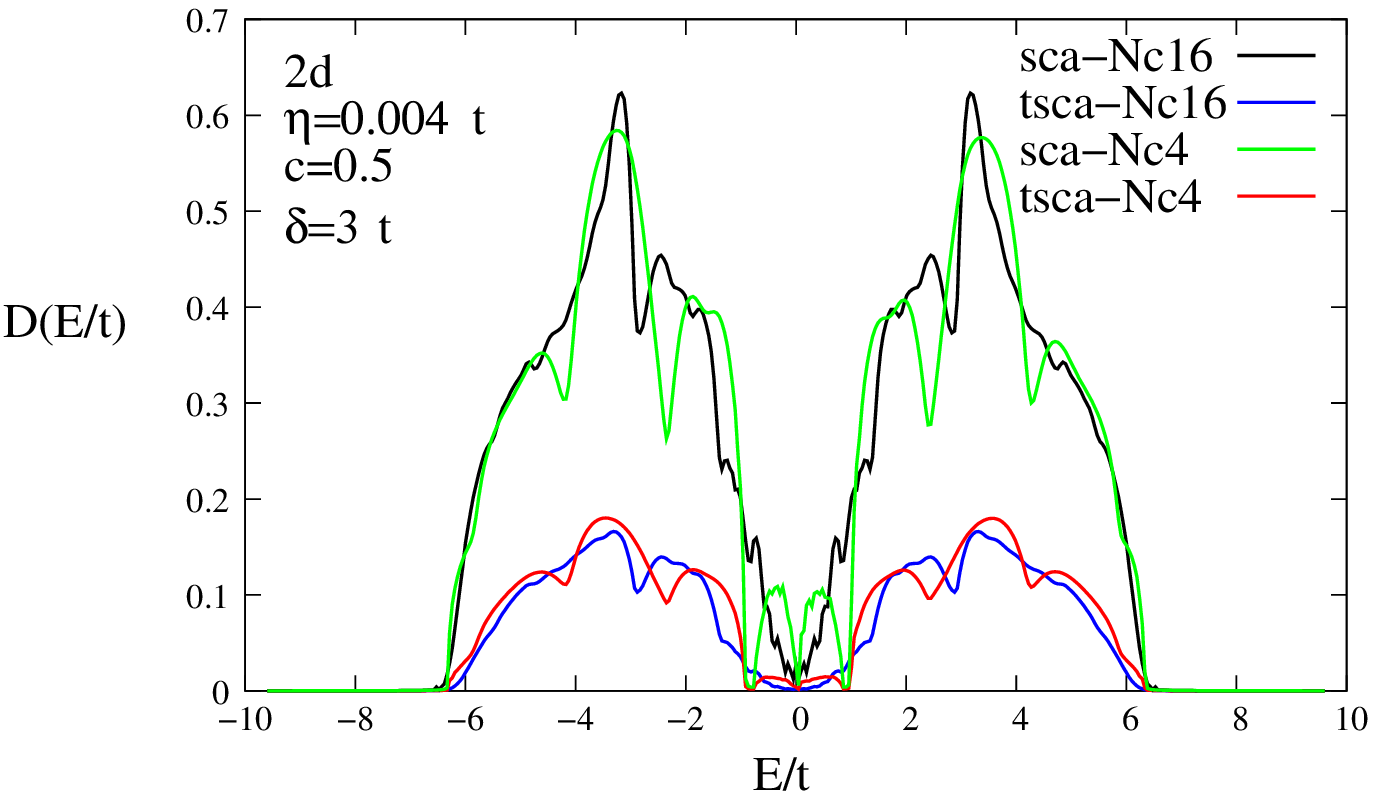  ,width=8.0cm,angle=0}}
	\caption{(Color on line) Shows sca, bsca and typical density of states for a two dimensional square lattice for different impurity strengths.  At strong strengths,  band splitting occurs and typical density of states at Fermi energy goes to zero hence electrons are localized.  }   
	\label{figure:localization-2d-typical} 
\end{figure}

 Fig.\ref{figure:localization-2d-typical} shows super-cell, beyond super-cell approximations density of states and localization order parameter that called typical density of states obtained by substitution super-cell self-consistent impurity Green function in the Eq.\ref{eq:typ-av-green-D}  for different impurity strengths.   
By increasing impurity strength, $\delta$, band splitting occurs and typical super-cell density of states at Fermi energy goes to zero that means all occupied states below Fermi energy separated from empty states which means electrons are localized. This results for a three dimensional alloy system observed by these authors\cite{Terletska14, Terletska17}.

\begin{figure}
	\centerline{\epsfig{file=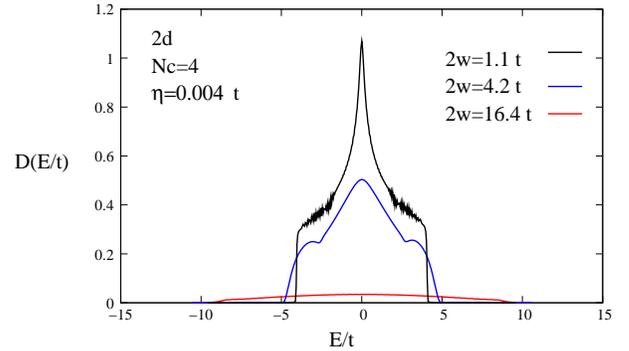   ,width=8.0cm,angle=0}}	
	\caption{(Color on line) Shows sca typical density of states for a two dimensional square lattice for different impurity strengths of box uniform model.  By increasing $W$ typical density of states decreases.  }   
	\label{figure:localization-2d-typical-uniform} 
\end{figure}

To complete our discussions we applied our methods to a square two dimensional system with uniform box impurity distribution where at $L$th site in the super-cell impurity probability distribution is $p(\varepsilon_{L})=\frac{1}{2W}\Theta(W-|\varepsilon_{L}|)$. The super-cell impurity average of impurity random function $X$ is given by $\bar{ X}=\Pi^{Nc}_{L=1} \int^{W}_{-W}d\varepsilon_{L}p(\varepsilon_{L}) X$. Fig.\ref{figure:localization-2d-typical-uniform} shows typical super-cell density of states for a square lattice for $2W=1.1 t, 4.2 t$ and $16.4 t$. By increasing disorder strength, $W$, typical density of states decreases and at strong $W$ it is goes to zero that means electron localization.  This result is same as Fig.2 of Ref.13 that observed by some of these authors. 

In summary we clearly showed that main part of Zhang et al. method\cite{Zhang} is a simple case of our super-cell approximation by ignoring $\{{\bf K}_{1},...,{\bf K}_{N_{c}}\}$ dependent of super-cell self-energies $\Sigma_{sc}({\bf K}_{n}; E)$ and replacing them by a local self-energy, $\Sigma_{sc}({\bf K}_{n}; E)=\Sigma_{sc}(L,L; E)$, which we called super-cell local self-energy approximation. In contrast to our super-cell approximation\cite{Moradian03} that the average Green function recovers exact average Green function in the limit $N_{c}\;\longrightarrow\;N$, the super-cell local self-energy approximation  average Green function converts to  $\bar{G}({\bf k}, E)=\frac{1}{E+i\eta+\epsilon_{\bf k}-\Sigma(l,l, E)}$ which does not recover exact average Green function, $\bar{G}({\bf k}, E)=\frac{1}{E+i\eta+\epsilon_{\bf k}-\Sigma({\bf k}, E)}$. Another big weakness point of their method is neglecting self-energy k-dependent that for low dimensional systems is strongly k-dependent.  
 Our method SCA systematically capture electron localization in contrast to DCA\cite{Jarrell} which is not able to show electron localization in substitution disorder systems. Since super-cell  approximation is able to show electron localization, in the  localization order parameter calculations using self-consistency of typical Green function is not necessary.

\end{document}